# Quantifying the Damage Induced by XPS Depth Profiling of Organic Conjugated Polymers

*Yvonne J. Hofstetter and Yana Vaynzof\**

Kirchhoff-Institut für Physik, Ruprecht-Karls Universität Heidelberg, Im Neuenheimer Feld 227, 69120 Heidelberg, Germany. Centre for Advanced Materials, Ruprecht-Karls Universität Heidelberg, Im Neuenheimer Feld 225, 69120 Heidelberg, Germany.

ABSTRACT

X-ray photoemission spectroscopy (XPS) depth profiling using monoatomic $Ar^+$ ion etching sources is a common technique that allows for the probing of the vertical compositional profiles of a wide range of materials. In polymer-based organic photovoltaic devices, it is commonly used to study compositional variations across the interfaces of the organic active layer with charge extraction layers or electrodes, as well as the vertical phase separation within the bulk-heterojunction active layer. It is generally considered that the damage induced by the etching of organic layers is limited to the very top surface, such that the XPS signal (acquired from the top ~10 nm of the layer) remains largely unaffected, allowing for a reliable measurement of the sample composition throughout the depth profile. Herein, we investigate a range of conjugated polymers and quantify the depth of the damage induced by monoatomic etching for $Ar^+$ ion energies ranging from 0.5 keV to 4 keV using argon gas cluster ion beam depth profiling. The results demonstrate that even when etching with the lowest available monoatomic ion energy for as little as 3 s, the damaged polymer material extends deeper into the bulk than the XPS probing depth. We show that the damaged material distorts the compositional information obtained by XPS, resulting in erroneous depth profiles. Furthermore, we propose that only gas cluster ion beam etching sources should be used for depth profiling of organic conjugated polymers, as those induce significantly less damage and maintain the compositional information throughout the entire profile.

INTRODUCTION

Over the past two decades, significant advances in the field of organic photovoltaic (OPV) devices have resulted in a steady increase in power conversion efficiencies, which have reached 14.2% and 17.3% for single and double junction PVs, respectively.[1-2] The active layer of these devices consists of a bulk-heterojunction (BHJ) blend of organic donor and acceptor materials. While spin-coating is a well-established method for the production of BHJs, there are still issues that need to be addressed in order to improve both device performance and reproducibility. For example, during film processing, de-mixing and crystallization heavily affect the blend's nano- and microstructure.[3-5] The resulting domain sizes influence fundamental processes such as exciton and charge transport,[6-7] while the surface compositions determine interfacial properties, such as charge separation efficiency and recombination.[8] Therefore, understanding the bulk and surface morphologies and composition is a key factor for controlling and improving the performance of OPVs.

X-ray photoemission spectroscopy (XPS) probes the elemental composition and chemical environment of the surface of a film, specifically the top 5-10 nm. When combined with sputtering and etching techniques, it can be used to create depth profiles and access this information for both the bulk of a material and buried interfaces of a device, making it an essential tool for the study of OPVs. Numerous organic material systems, including both full devices and semiconductor films, have been investigated using XPS depth profiling, wherein small projectiles, such as argon ions, most commonly make up the sputtering beam. For example, we and others have used this method to study phase separation in poly(3-hexylthiophen-2,5-diyl) (P3HT) and [6,6]-phenyl-C61-butyric acid methyl ester (PCBM) BHJ films, which form a P3HT-rich layer at the air interface and a PCBM-rich layer at the bottom interface.[9-13] Similarly, we found that P3HT and poly[(9,9-dioctyluorene)-2,7-diyl-alt-(4,7-bis(3-hexylthien-5-yl)-2,1,3-benzothiadiazole)-2',2''-diyl] (F8TBT) BHJs exhibit an enriched P3HT top layer due to its lower surface energy.[14] XPS depth profiling has also been applied to study vertical phase separation in other BHJ organic systems[15] and ternary blends,[16] as well as effects of interfacial modifications.[17] Furthermore, XPS depth profiles can provide insight into degradation mechanisms in organic electronic devices, thereby tackling the issue of long-term stability. In one example, Dupont *et al.* investigated moisture-assisted decohesion of poly(3,4-ethylenedioxythiophene) polystyrene sulfonate (PEDOT:PSS) films by tracking the oxygen concentration throughout the film by XPS.[18] In another, Kumar *et al.* monitored degradation and regeneration processes of P3HT:PCBM BHJ solar cells revealing the chemical reactions taking place at the active layer/cathode interface.[19] These examples are only a small sample of the numerous studies in which XPS depth profiling has been applied to the study of organic materials and organic electronic devices. Furthermore, this technique has also been applied to the study of hybrid organic-inorganic devices[20-22] and most recently to lead halide perovskite devices[23-26] - all of which contain organic materials as part of the device structure.

However, the validity of the XPS depth profile result relies on the assumption that any damage induced by the etching process is limited to the very top surface of the remaining material, such that due to its relatively high probing depth (~10 nm), the XPS signal originates predominantly from deeper, undamaged material. While this has been shown to be the case for many inorganic materials,[27] this is not necessarily the case for soft organic materials. It is important to keep in mind that non-negligible damage may be introduced to the surface as well as the underlying organic layers upon ion impact. This damage may take many forms, for example: molecular fragmentation, reduction, cross-linking or preferential etching. The formation of surface damage is especially critical when trying to extract exact compositional profiles and chemical states from the measurement. It is thus necessary to quantify the damage induced by monoatomic etching and compare it to the probing depth of XPS (~10 nm). Despite the common use of XPS depth profiling in organic electronics research, the damage depth of monoatomic etching has never been quantified for any organic conjugated polymers.

One possible method to quantify the depth of damaged material is based on the use of an alternative, non-detrimental sputtering technique utilizing gas cluster ion beams (GCIB) or molecular clusters. Such gentler sputtering techniques were first explored in combination with secondary ion mass spectrometry (SIMS),[28-32] but more recently have also been employed in combination with XPS,[33-34] allowing for an essentially damage free depth profiling of organic materials.[35-36] For example, in the case of PEDOT:PSS films, Yun *et al.* have shown that while monoatomic $Ar^+$ etching alters the bonding states of C, S and O, changing the chemical composition of the bulk material, argon cluster etching preserves the PEDOT:PSS core level structure.[37]

To quantify the damage induced by a monoatomic $Ar^+$ beam, one strategy is to first etch for a set period of time using a monoatomic beam, and then follow this step with a depth profile using an argon GCIB until the original material compositional structure is recovered. Such an approach was used by Miyayama *et al.,* who reported that polyimide (PI) surfaces exhibited damaged N species, and reduced N and O content upon monoatomic $Ar^+$ ion etching, which could subsequently be removed using an argon GCIB of 10 keV, with only 17% damaged N species remaining.[38] The dose required to remove the damaged layers scaled linearly with the monoatomic $Ar^+$ beam energy. Similarly, Yancey and coworkers observed a reduction of O and F in poly(ethylene terephthalate) (PET) and polytetrafluoroethylene (PTFE), respectively, upon monoatomic $Ar^+$ ion sputtering with the damaged layers being entirely removed using an argon GCIB.[27]

Herein, we investigate and quantify the damage induced by monoatomic $Ar^+$ sputtering on conjugated polymers, such as those used in organic electronic devices. We focus on four polymers which are frequently used in organic electronic devices: regioregular and regiorandom P3HT (c-P3HT and a-P3HT, respectively), poly[(9,9-di-n-octylfluorenyl-2,7-diyl)-alt-(benzo[2,1,3]thiadiazol-4,8-diyl)] (F8BT), poly({4,8-bis[(2-ethylhexyl)oxy]benzo[1,2-b:4,5-b']dithiophene-2,6-diyl}{3-fluoro-2-[(2-ethylhexyl)carbonyl]thieno[3,4-b]thiophenediyl}) (PTB7), and poly[4,8-bis(5-(2-ethylhexyl)thiophen-2-yl)benzo[1,2-b;4,5-b']dithiophene-2,6-diyl-alt-(4-(2-ethylhexyl)-3-fluorothieno[3,4-b]thiophene-)-2-carboxylate-2-6-diyl)] (PCE10). We apply an argon GCIB to perform XPS depth profiles on pristine films, as well as films that have been previously etched for 3 s with a monoatomic $Ar^+$ beam. The depth profiles of pristine films show no significant changes, confirming that the GCIB is indeed non-detrimental to the investigated polymers. However, 3 s of monoatomic etching is enough to change the depth profiles drastically, revealing the detrimental effect of the argon ions on these polymers, observed in the form of XPS peak broadening, broken bonds and altered atomic ratios. We successfully remove the layers that were damaged by the monoatomic etching with cluster etching and gain quantitative information about the damage depth of monoatomic etching. We find that the damage depth scales linearly with the monoatomic etch energy while the exact magnitude is material but not crystallinity dependent. Most importantly, we find that the damage depth is larger than the XPS probing depth even for the lowest possible monoatomic $Ar^+$ ion energy of 0.5 keV, indicating that XPS depth profiling based on monoatomic $Ar^+$ is not suitable for conjugated polymers and the data must be treated carefully. We hope our work will motivate researchers to avoid using monoatomic etching for XPS depth profiling of organic layers or organic electronic devices and provide guidelines for future analysis of such studies.

EXPERIMENTAL METHODS

**Materials** Regiorandom P3HT (RR < 80), regioregular P3HT (RR > 96), PTB7 and PCE10 were purchased from 1-Material, and F8BT was purchased from Ossila. Their chemical structures are shown in Figure 1. All other chemicals were purchased from Sigma-Aldrich. All materials were used as received.

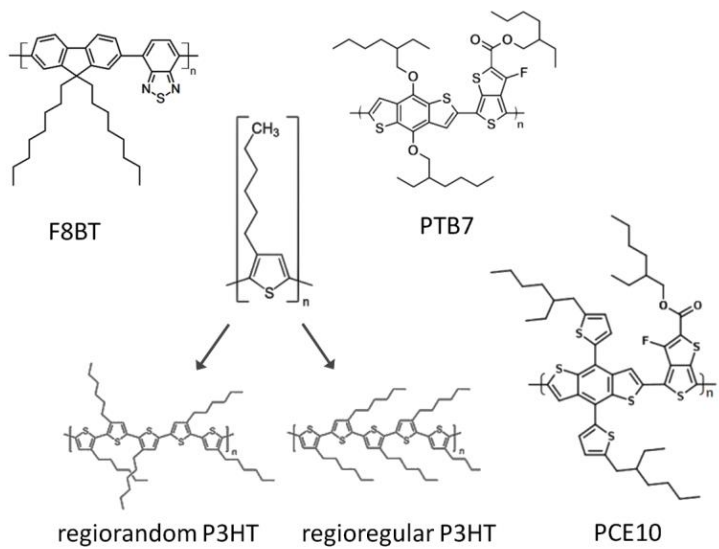

**Figure 1.** Chemical structures of the investigated polymers.

**Sample preparation** The samples were prepared in ambient conditions. First, ITO-coated glass sheets (MSE Supplies) were cut into (20x20) mm² substrates and then subsequently sonicated in acetone and isopropanol for 5 min each, followed by a 10 min oxygen plasma treatment. Afterwards, the zinc oxide (ZnO) sol-gel was spin-coated at 2000 rpm for 45 s and annealed at 200 °C for 30 min.[39-40] Next, the polymer solutions were prepared in concentrations ranging from 15 mg/ml to 50 mg/ml depending on the material; P3HT and PCE10 were dissolved in chlorobenzene, PTB7 in chloroform, and F8BT in toluene. The polymer solutions were spin-coated on top of the ZnO layer at 1000 rpm to 2000 rpm for 45 s depending on the polymer. Optionally, the polymer films were annealed for 10 min; P3HT and PCE10 at 140 °C, PTB7 at 80 °C, and F8BT at 155 °C.

**X-Ray Photoemission Spectroscopy (XPS) and Etching** The samples were transferred to an ultrahigh vacuum chamber (ESCALAB 250Xi by Thermo Scientific) for XPS measurements and etching. XPS measurements were carried out using a XR6 monochromated Al Kα source (hν = 1486.6 eV), a pass energy of 20 eV and an oval measurement spot with a long axis of 900 μm.

Etching was carried out with the MAGCIS dual mode ion source, which can be operated as a monoatomic argon ion source or as an argon gas cluster ion source. The monoatomic mode was operated at energies ranging from 0.5 keV to 4 keV, and the cluster mode always at 4 keV with 'large' clusters ($Ar_{2000}^+$). The etched spot was (2x2) mm$^2$ in size.

**Measurement procedure** The measurement procedure is schematically depicted in Figure 2. First, a spot on the surface of the polymer film was etched with the monoatomic $Ar^+$ beam for 3 s, creating damage within the film. This short duration of etching is insufficient to remove any substantial amount of material from the film surface, but already induces damage in the film. Next, the spot was depth-profiled with the argon cluster beam (4 keV) until the polymer/ZnO interface was reached. The time of the cluster etch steps was adapted to each material individually to achieve a depth resolution of roughly 3 nm for the top 50 nm of the film. Afterwards, the resolution was decreased so that etching would be faster while approaching the interface. XPS spectra were recorded in between each etching step. The reference depth profile (cluster 4 keV) follows the same procedure but without the 3 s of monoatomic etching. We determined the etch time to etch depth conversion by evaluating the time needed to reach the polymer/ZnO interface, and measuring the etched polymer film thickness with a DektakXT profilometer (Bruker). Usually, the damage from monoatomic source etching was tracked by an increased full width at half maximum (FWHM) of the XPS peaks. The choice of monitoring the evolution of the FWHM, rather than introducing new peaks in order to fit the spectra of the damaged layers, was motivated by the fact that the latter procedure would entail speculative decisions about the number of peaks that are associated with damaged chemical species and their relative ratios, and would not provide definitive information about the species formed as a result of the etching. Therefore, the FWHM over-etch depth was plotted in comparison to the cluster-only reference measurement to extract the monoatomic source etching induced damage depth. The measurement uncertainties in film thickness and etch time generally resulted in an etch depth error of 5-10%. In total, five monoatomic etch energies were tested: 0.5 keV, 1 keV, 2 keV, 3 keV, and 4 keV. All spots that were etched with the monoatomic source and the reference spot were located on the same sample.

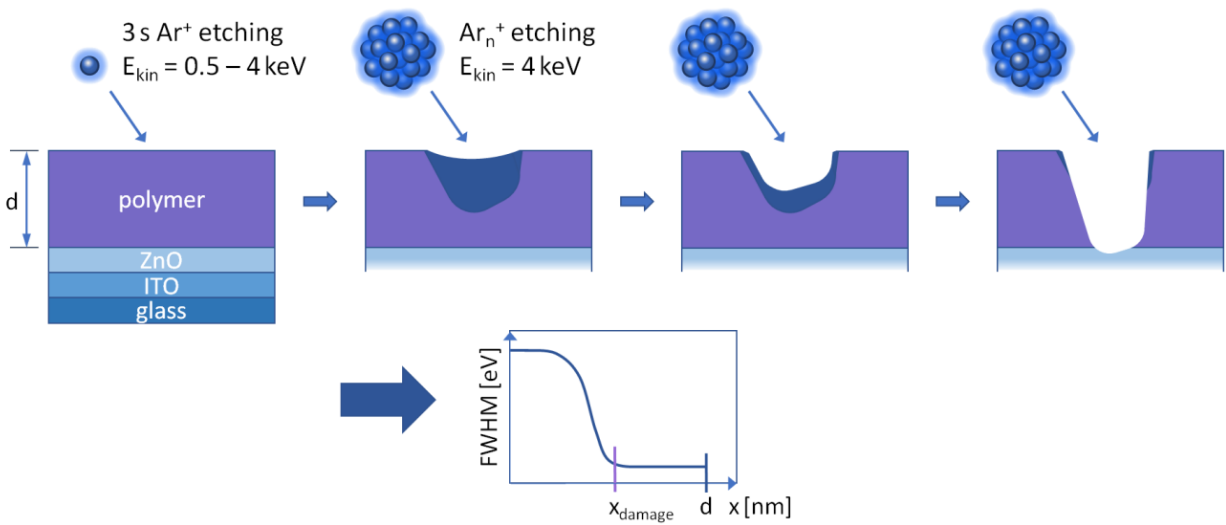

**Figure 2.** Schematic of the experimental procedure. First, the polymer surface is etched with a monoatomic argon ion beam for 3 s creating damage at the surface and the underlying layers of the polymer. Then the polymer film is depth profiled using an argon GCIB until the polymer/ZnO interface is reached. From the resulting FWHM depth profile, the damage depth $x_{damage}$ can be extracted.

RESULTS AND DISCUSSION

**Quantification of the damage depth**

P3HT is one of the most investigated organic conjugated polymers, with applications in a range of optoelectronic devices.[41] We begin by investigating the damage induced in amorphous P3HT (a-P3HT) by argon etching. Prior to etching, the FWHM of the S $2p_{3/2}$ and S $2p_{1/2}$ peaks is in the range 0.80 to 0.85 eV. Upon cluster etching at 4 keV, the FWHM increases slightly, to just under 0.90 eV. The shapes of the S 2p doublet and the FWHM are preserved throughout the profile, even upon cluster etching for longer than 2000 s. The S 2p peak evolution through 2000 s can be seen in Figure 3a. The S 2p peak evolution shown in Figure 3b starts with a reference surface measurement before etching. After this initial measurement, the spot is bombarded with a monoatomic beam for 3 s at 3 keV monoatomic $Ar^+$ ion energy, and measurement through argon cluster depth profiling continues as in Figure 3a. It is clear that the monoatomic etching causes significant changes: the S 2p peak is shifted to lower binding energies and appears broadened with a FWHM of up to 1.4 eV. Both changes indicate that the C-S bonds in the P3HT film have been damaged by the short monoatomic etching. Cluster depth profiling the spot that was etched using the monoatmic $Ar^+$ source gradually reduces the broadening and shift, eventually uncovering an S 2p peak shape like the one shown in Figure 3a. A more detailed evaluation of the FWHM reveals that it takes approximately 700 s to remove the damaged a-P3HT, corresponding to a depth of approximately 40 nm. In summary, these findings demonstrate that while 3 s of monoatomic etching barely removes any material from the surface, it is already sufficient to cause significant damage deep inside the film. Furthermore, cluster etching is capable of removing the material that has been damaged by monoatomic etching, allowing for accurate quantification of the damage depth.

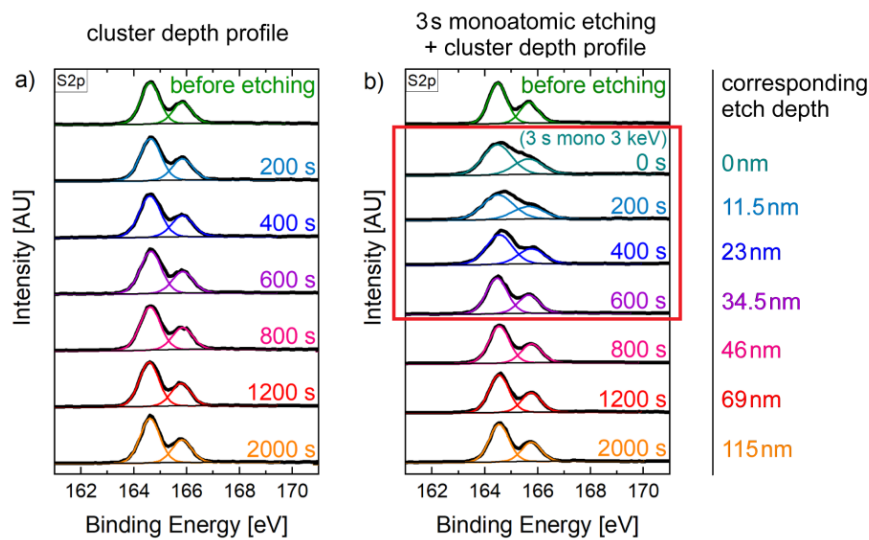

**Figure 3.** S 2p peak of an annealed a-P3HT film measured (a) before and during a cluster depth profile (4 keV) and (b) before and after 3 s of monoatomic etching (3 keV) followed by a cluster depth profile (4 keV); the spectra were aligned to the energy of the respective S $2p_{3/2}$ peak before etching. The red box marks the spectra damaged by monoatomic etching. Respective cluster etch times are given on the right-hand side of each spectrum.

To identify the damage depth for various ion energies, the experiments were repeated with monoatomic ion energies ranging from 0.5 keV to 4 keV, with the FWHM of the collected S 2p spectra shown as a function of depth in Figure 4a. The reference measurement shows that the FWHM increases slightly upon cluster etching and stabilizes just below 0.9 eV as the layer is gradually removed. In contrast, etching with the monoatomic beam for 3s increases the FWHM by up to 70% when compared to the initial value. Damage induced by a high monoatomic etch energy results in a larger increase in FWHM that reaches deeper into the film bulk. When all the damaged material has been removed by cluster etching, the FWHM is restored to the value of the reference measurement. The same trend can be seen in the C 1s spectra (Figure 4b). The point at which the FWHM returns to the level of the reference measurement was defined as the damage depth caused by monoatomic etching. The damage depth values extracted from the S 2p and the C 1s spectra are in good agreement with each other (Figure 4c).

It is interesting to compare the extracted damage depths to the Ar ion penetration depth estimated by the Stopping and Range of Ions in Matter (SRIM) program.[42] We find that the penetration depth varies from 9 nm for 1 keV ion energy to 20 nm for 4 keV Ar ions. These estimated penetration depths are substantially smaller than the measured damage depths, suggesting that much of the damage in the deeper layers is associated with propagation of free radicals that are formed during the etching process, rather than the Ar ions themselves.

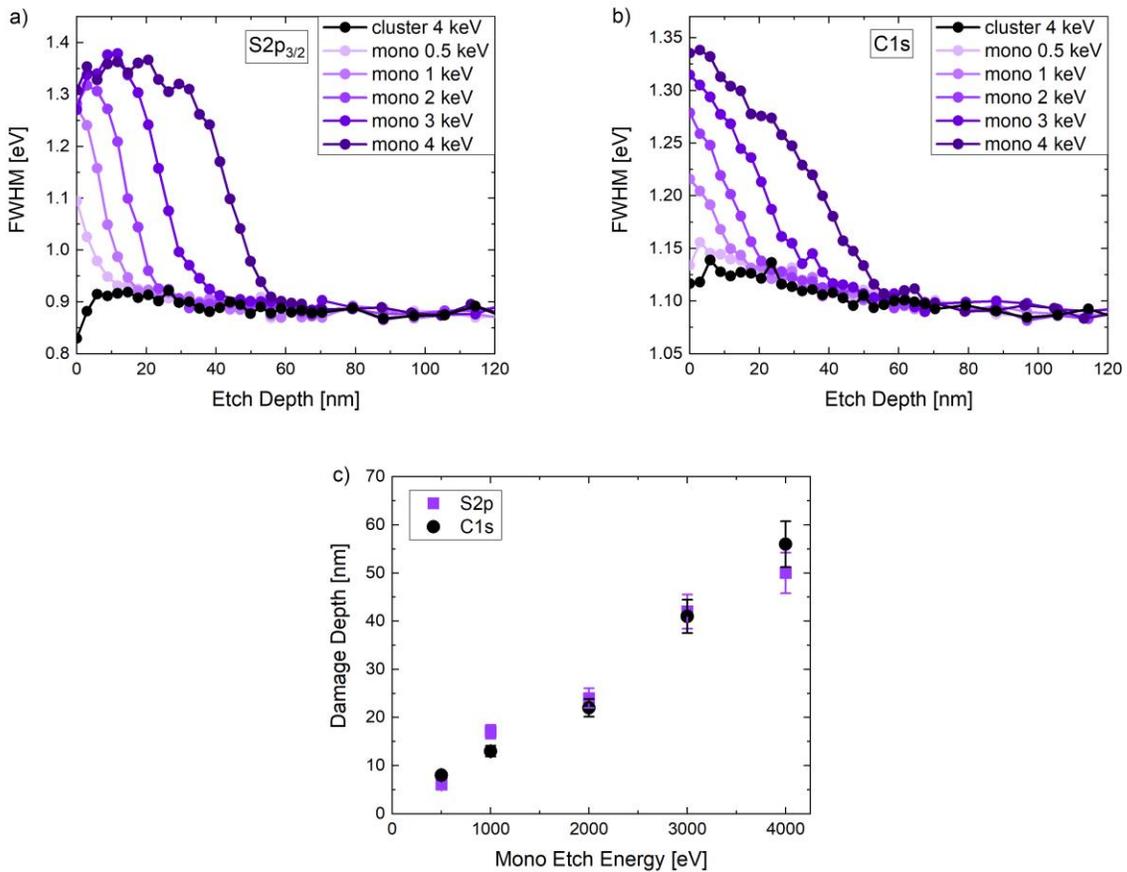

**Figure 4.** FWHM depth profiles of (a) S 2p and (b) C 1s measured on an annealed a-P3HT film. Shown is the cluster reference measurement and five measurement spots that have been exposed to 3 s of monoatomic etching of different energies. The damage depth with respect to monoatomic source energy obtained from the S 2p and C 1s FWHM profiles is shown in (c).

The conjugated polymers used in organic electronic devices show varying degrees of crystallinity depending on their chemical structure and processing parameters. To investigate whether material crystallinity plays any role in determining the depth of the induced damage, we repeated the experiment on samples of regioregular and regiorandom P3HT,[43] which were either measured as cast or annealed to induce different levels of crystallization. This allowed us to compare results from films with varying degrees of crystallinity with unannealed regiorandom P3HT (a-P3HT) being the most amorphous and annealed regioregular P3HT (c-P3HT) being the most crystalline film. The damage depth was then extracted from the respective S 2p FWHM depth profiles, as shown in Figure 3a for annealed a-P3HT and Figure S1 in the supplementary information (SI) for c-P3HT and unannealed a-P3HT. Figure 5 shows that the damage depth of all four P3HT samples scales linearly with the energy of the monoatomic Ar$^+$ source. The lowest monoatomic source etch energy, 0.5 keV, yields a damage depth of approximately 10 nm, while the highest monoatomic Ar energy, 4 keV, yields a damage depth of up to 60 nm. The results are comparable within the range of error for all four samples. Hence, the damage depth shows little to no dependence on the crystallinity of the conjugated polymer. It is important to note that the observed damage depth is equal to or deeper than the XPS probing depth ($\leq$ 10 nm) for all monoatomic Ar energies. In other words, after exposing P3HT to a monoatomic beam for only 3s even at low energies, all of the material contributing to the XPS signal is damaged.

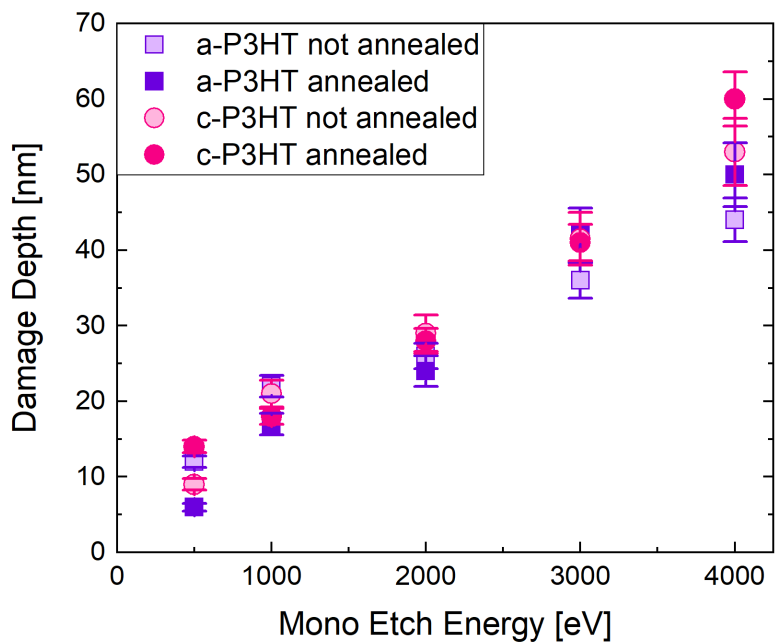

**Figure 5.** Damage depth of a-P3HT and c-P3HT in dependence on the monoatomic etch energy. The damage depth was extracted from the S2p FWHM.

We now examine the effect of etching on a polymer with a very different chemical structure - F8BT. While the damage induced upon etching of P3HT with a monoatomic $Ar^+$ source results solely in the increase of the FWHM of the S 2p and C 1s spectra, the damage induced in the S 2p peak of F8BT takes a different form. The evolution of the S 2p peak of F8BT (Figure 6a) upon cluster etching shows no significant changes; however, after very long etching times a small additional S 2p doublet appears at ~2.5 eV lower in binding energy than the main doublet. After etching for 100 nm, this lower peak accounts for 10% of the total S content (Figure S2a and b, SI), suggesting that cluster etching (4 keV) slightly damages the F8BT. However, after just 3 s of monoatomic etching (3 keV), the original S 2p peak has completely vanished and only the damaged S species can be observed (Figure 6b). After again employing GCIB etching, the damaged species is removed and the main S peak reappears. Eventually, the percentage of the damaged S species declines to the level of the reference measurement marking the damage depth. The percentage of damaged S species is shown in Figure S2a and b in the SI. The main peaks of the other elements in F8BT, C and N, appear broadened after monoatomic Ar$^+$ etching, but do not show any new peaks.

These results suggest that the chemical structure of the polymer strongly affects the type and degree of damage induced by the monoatomic etching. In the case of P3HT, the thiophene ring is more robust to etching damage, while the benzothiadiazole (BT) unit in F8BT is weaker, resulting in a higher degree of damage. This is consistent with the bond strength differences between the C-S in thiophene (713.3 kJ/mol) and the N-S in BT units (467 kJ/mol) due to dissimilarities in electronegativity.[44] This is not only supported by the spectral changes, but also by the quantification of the damage depth, which consistently shows larger damage depths for F8BT when compared to P3HT. Since many organic polymers of interest in organic photovoltaics contain both types of units,[45-46] extracting quantitative compositional information from XPS depth profiling of these polymers should be performed with extreme care when using monoatomic etching.

Even when the polymers consist of similar chemical building blocks, the exact chemical structure may influence the damage depth. For example, two high photovoltaic performance conjugated polymers used in photovoltaic devices, PTB7 and PCE10, consist of the same types of S-containing chemical moiety: thiophene and fluorothieno[3,4-b]thiophene. In both cases, the FWHM of the S 2p peak increases upon etching and was used as the damage indicator. The FWHM depth profiles of PTB7 and PCE10 (Figure S2c-f, SI) show that induced damage is consistently deeper in PCE10 than in PTB7 regardless of monoatomic ion energy. This is likely to be a result of the inclusion of the two thiophene rings at the central ring of the benzodithiophene core, instead of the alkoxy chains in PCE10. The former can probably be easily cleaved by monoatomic etching and result in the deeper observed damage.

Figure 6c summarizes the damage depths of all annealed polymer films. A linear trend of increasing damage depths with increasing monoatomic etch energy is prominent for all five materials. However, the slopes are different. F8BT overall shows a slightly deeper damage than the P3HT, reaching over 60 nm upon monoatomic $Ar^+$ etching at 4 keV energy, whereas PTB7 and PCE10 are less prone to damage with their highest damage depths being around 28 nm and 39 nm for the same monoatomic ion energy ions (4 keV), respectively. For all investigated materials, no significant difference between annealed and unannealed samples was observed (Figure S3, SI), reinforcing the previously mentioned independence of damage depth on polymer crystallinity or microstructure. This suggest that the type and extend of damage introduced by monoatomic $Ar^+$ etching is dependend on the individual molecular structure and electronic properties of the polmers rather than on their solid state packing and interchain interactions.

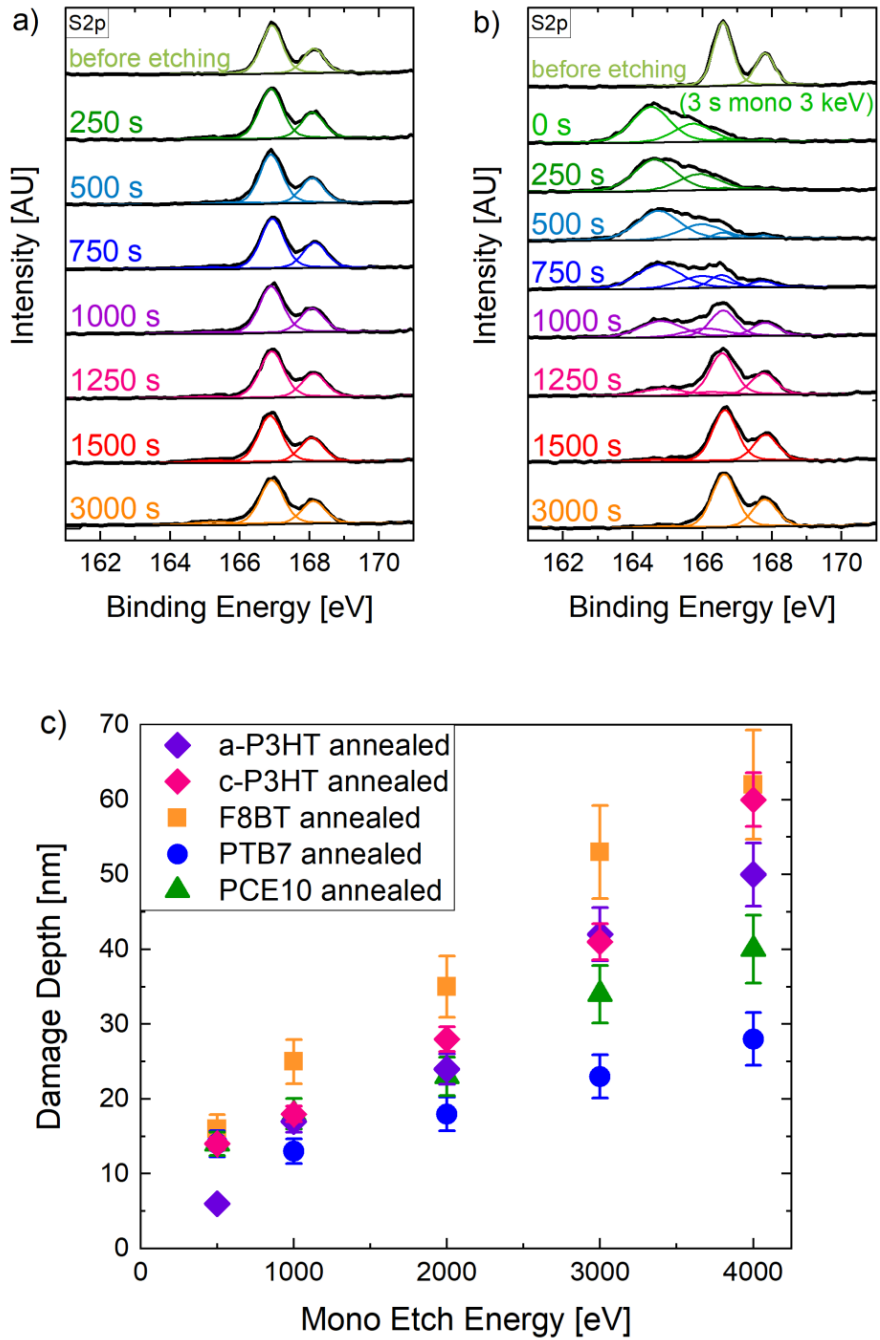

**Figure 6.** S 2p peak of an annealed F8BT film measured (a) before and during a cluster depth profile (4 keV) and (b) before and after 3 s of monoatomic etching (3 keV) followed by a cluster depth profile (4 keV); the spectra were aligned to the energy of the respective S $2p_{3/2}$ peak before

etching. The damage depth of F8BT, PTB7, and PCE10 with respect to the monoatomic etch energy, in addition to the previously shown a-P3HT and c-P3HT, are shown in (c).

**Effect of etching with a monoatomic Ar$^+$ source on the extracted compositional profiles**

The observed changes in the spectral shapes of the elemental constituents of the polymers are tolerable in the context of XPS depth profiling if they do not affect the extracted atomic compositional information obtained by XPS. To investigate the effects of monoatomic Ar$^+$ etching on the extracted composition of the polymers, we compare the measured atomic composition to that which is expected from the chemical structure of the polymer. Figure 7 shows a depth profile of PCE10 focusing on S, O, and F. Not shown is the contribution of C, which makes up the rest of the polymer. The cluster reference measurement shows constant values for all elements of PCE10 throughout the film upon etching, with reasonably good agreement with the theoretical values. We note that the surface measurement shows excess O due to surface contamination, which is removed with the first etching step. However, all samples that have been subjected to 3 s of monoatomic etching show a significant reduction in O and F content in the damaged layers, while the effect on S is less pronounced. Correspondingly, the relative C content increases, complementing the decrease of O and F. These results suggest that bonds containing O and F are particularly prone to damage upon monoatomic Ar$^+$ bombardment. Similar observations can be made in the case of PTB7, which also consists of S, O, F and C, just like its derivative PCE10. A strong reduction in F and O after monoatomic etching is observed as well (Figure S4c, SI), while the C and S content is complementary higher. In the case of F8BT, where the altered peak position indicates that an S-N bond of the thiadiazole ring must have been damaged, a reduction in N was observed. However, the total percentage of damaged and undamaged S is constant throughout the film (Figure S4b, SI). These observations illustrate that monoatomic etching can preferentially attack certain chemical bonds, and can therefore significantly alter the composition of polymer films. These results suggest that bonds containing electronegative atoms such as F, O or N are particularly susceptible to damage when bombarded with monoatomic $Ar^+$ ions. It is possible that these bonds are cleaved or weakened, such that even upon etching with cluster argon source, they are preferentially removed, leading to a misrepresentation of the original composition of the polymer. This is of critical importance to XPS depth profiling in organic electronics, where accurate compositional information is used for the study of materials or interpretation of device properties. For example, in the case of depth profiling BHJs of a donor polymer and a small molecule acceptor (typically a fullerene), a specific element (commonly not carbon) must be chosen to track each component throughout the depth profile. Our results show that the choice of different elements within the polymer will strongly influence the obtained compositional profile.

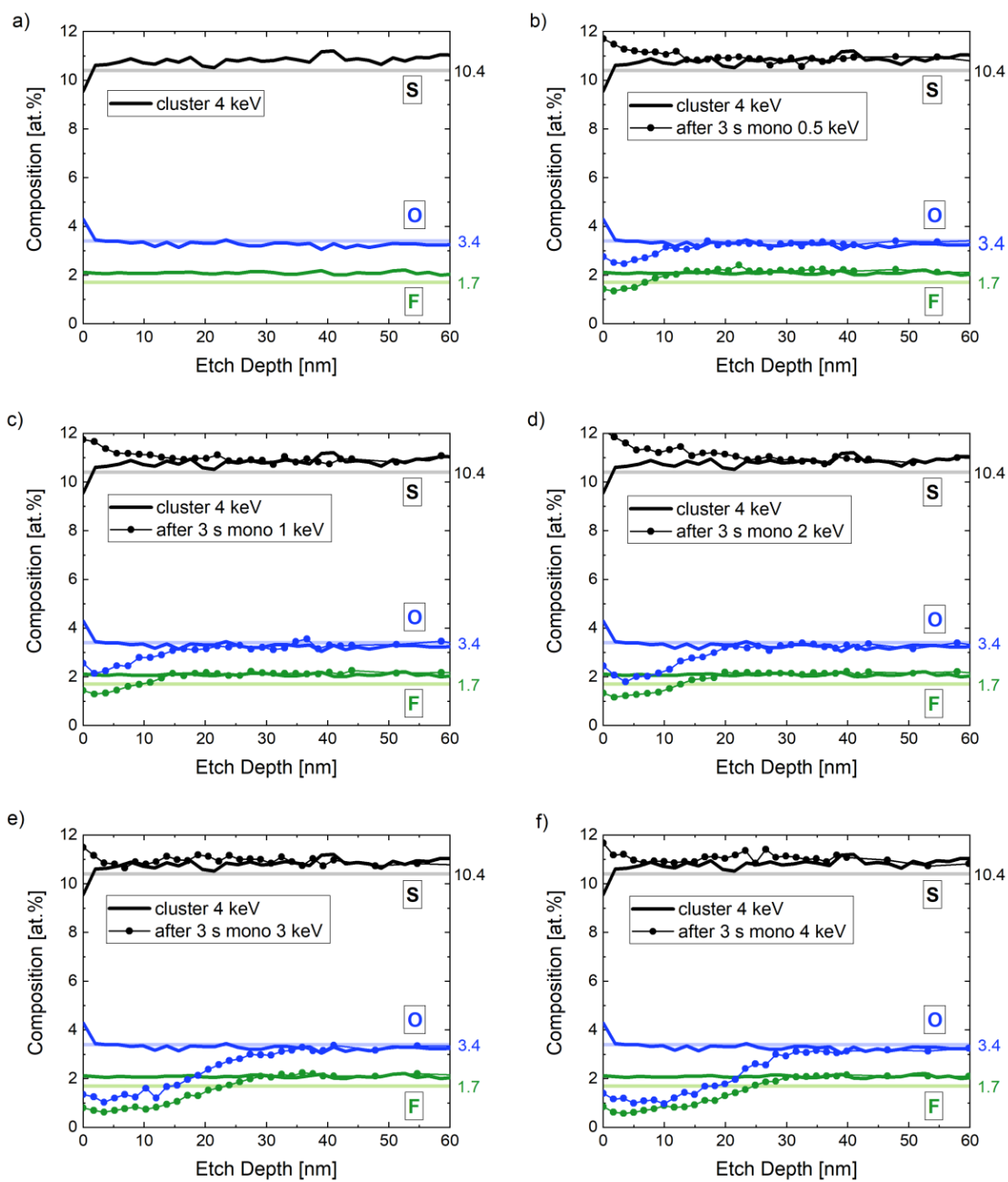

**Figure 7.** Compositional depth profiles of PCE10 showing the S, O, and F percentage of (a) the cluster reference measurement and after exposure to 3 s of (b) 0.5 keV, (c) 1 keV, (d) 2 keV, (e) 3 keV, and (f) 4 keV monoatomic etching. The theoretically expected atomic percentages of each element are given on the right-hand side and as straight lines in the graphs.

**Effect of monoatomic Ar$^+$ etching time**

It is noteworthy that for P3HT, no change in atomic composition is observed upon monoatomic etching for 3 s (Figure S4a, SI). This might suggest that some polymers are resilient enough to be accurately studied by XPS depth profiling using monoatomic etching sources. However, it is important to remember that etching for 3 s is not sufficient to achieve any substantial material removal, and longer etching times are required for characterizing the bulk of materials. To investigate whether the damage induced by monoatomic etching is confined to a certain depth, regardless of etching time, we varied the monoatomic etching time from 3 s to 300 s (5 min). As before, we then characterized the extent of the damage by removal of the damaged material using the GCIB (4 keV), until the measured XPS spectra matched the reference. We chose to investigate P3HT as it is the only material that shows unaltered atomic composition for 3 s of monoatomic Ar$^+$ etching.

Figure 8a shows the S 2p FWHM depth profile of an annealed a-P3HT film etched with a 3 keV monoatomic beam for varying durations in different spots. The etch depth has been corrected for the material etched away by the monoatomic beam estimated via the etch rate determined for monoatomic etching at 3 keV (0.044 nm/s). The damage depth after 3 s of monoatomic etching is comparable to the previous results (Figure 4c and 5), and increases with longer etch times until it seemingly saturates after several minutes. For 30 s and longer, the top 100-150 nm of the film is heavily damaged, with decreasing damage deeper in the film. However, the FWHM cannot be restored to the level of the reference measurement within this 300 nm thick film. Therefore, we conclude that the after 30 s of monoatomic etching at 3keV the entire film is damaged. Figure 8b shows the corresponding compositional depth profile, where no significant changes are present for 3 and 10 s of etching with a monoatomic Ar$^+$ source. For 30 s and longer, the composition is altered in the top 100 to 150 nm of the film, corresponding to the heavily damaged region seen in the FWHM depth profile. In this region, the S content is heavily reduced by up to 30%, and will result in a significant underestimation of P3HT content, if S is used as an indicator for P3HT. We note that the etching energy of 3 keV has been chosen as it appears to be very common in literature studies that report XPS depth profiling of organic layers.[13,15-16,19-21] Repeating the experiment at a lower monoatomic etch energy of 1 keV revealed that longer etch times also lead to deeper damage in the form of an increased FWHM and a reduced S content (see SI, Figure S5).

These results demonstrate that the damage induced by monoatomic etching accumulates as the etching time increases. We note that etching for 30 s is a realistic etch step size for depth profiling P3HT, resulting in a depth resolution of 1.3 nm for a 3 keV beam in our case. However, the depth profile obtained will simply be an artifact, since the entirety of the polymer film will have been damaged during the first step. In the case of the 1 keV beam, the damage depth for a 30 s etch step is approximately 35 nm, far higher than the XPS probing depth, rendering the obtained XPS depth profiles equally untrustworthy. For both monoatomic ion energies, the damage depth is far larger than the estimated penetration depth of Ar ions, again pointing out to propagation of free radicals as the most likely cause of damage at the deeper layers.

We also note that prolonged etching with a monoatomic $Ar^+$ beam results in significant cross-linking of the polymer. This cross-linking results in a reduction in the etching rate, resulting in an increase in the overall etching time required to reach the interface with the ZnO substrate. While for short monoatomic etching time of 3 s that resulted in a damage of 40 nm, the overall etching time remained largely unchanged (within 5-10%), the etching time for a film exposure to monoatomic beam for 300 s has increased by 260%. This significant change in etch rate may result in preferential etching when monoatomic etching is applied to a blend of another component with P3HT, adding uncertainty to the results of XPS depth profiling experiments. Taking these results together suggests that XPS depth profiling using monoatomic etching is unsuitable for application on organic conjugated polymers.

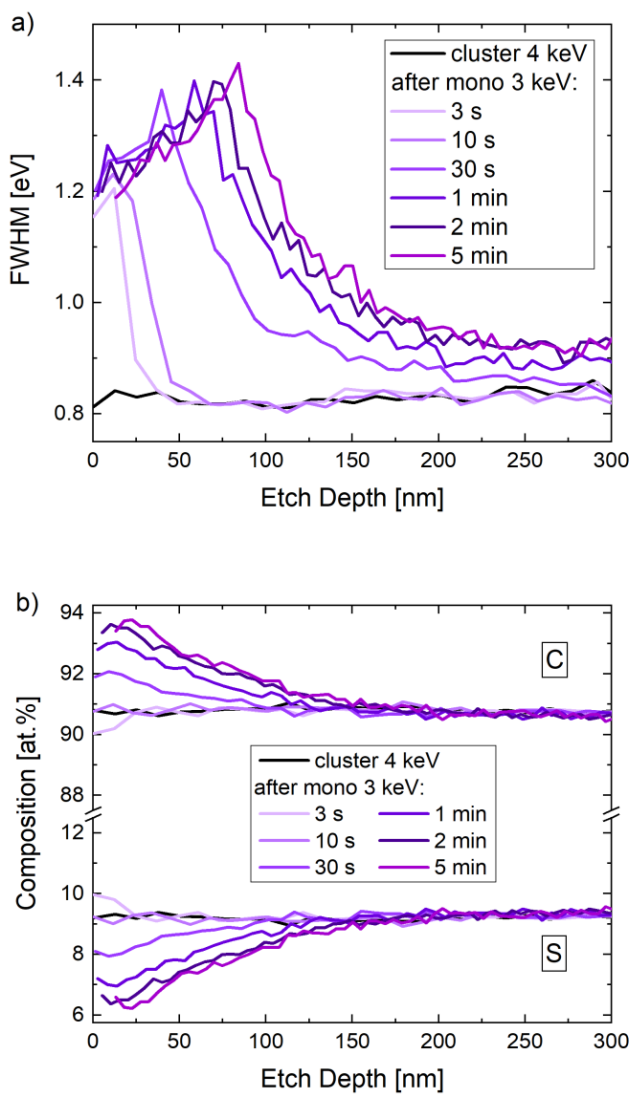

**Figure 8.** (a) S 2p FWHM and (b) compositional depth profiles of an annealed a-P3HT film after monoatomic etching (3 keV) of varying duration. The depth profiles have been corrected for the material removed by the monoatomic etching, which amounted to 0.1 to 13 nm depending on the mono etch time.

CONCLUSIONS

In this work, we studied the effect of argon etching on four different polymers: P3HT (regiorandom and regioregular), F8BT, PTB7 and PCE10 using monoatomic and gas cluster beams. XPS measurements show that there are no substantial changes in any film upon argon cluster etching (4 keV), whereas only 3 s of monoatomic argon etching on the surface is enough to cause severe damage within the polymer films, generally observed as peak broadening or, in the case of sulfur in F8BT, as a damaged species. Additionally, the atomic ratios were altered for all polymers except P3HT. We successfully employed cluster etching to remove the damaged layers, revealing a damage depth of approximately 10 to 60 nm, which varied linearly depending on the ion energy. The damage depth was consistent for both FWHM and composition. It was found to be independent of the film crystallinity, but material specific. Measurements on a-P3HT showed that damage accumulates when the mono-etch time is increased to 30 s or longer - even altering the previously resistant composition of P3HT up to a depth of 150 nm. This is far beyond the XPS probing depth of ≤ 10 nm, meaning that any XPS depth profiles carried out using monoatomic etch beams will only measure damaged polymer layers and should be treated with caution. Therefore, gas cluster etching is mandatory in order to obtain reliable results - especially when composition plays a crucial role.

## ASSOCIATED CONTENT

**Supporting Information**.

The following files are available free of charge. FWHM depth profiles of c-P3HT (annealed, unannealed) and a-P3HT (unannealed); Depth profiles of the damaged S species in F8BT and FWHM depth profiles of PTB7 and PCE10 (all annealed, unannealed); Compositional depth profiles of a-P3HT, F8BT, and PTB7 (all annealed); FWHM and compositional depth profiles of c-P3HT (annealed) after monoatomic etching (1 keV) of varying duration. (PDF)

## AUTHOR INFORMATION

**Corresponding Author**

*E-mail: vaynzof@uni-heidelberg.de

**Author Contributions**

The manuscript was written through contributions of all authors. All authors have given approval to the final version of the manuscript.

**Funding Sources**

This project has received funding from the European Research Council (ERC) under the European Union's Horizon 2020 research and innovation programme (ERC Grant Agreement n° 714067, ENERGYMAPS).

## ACKNOWLEDGMENT

The authors would like to kindly thank Prof. U. Bunz for providing access to film fabrication facilities.